\documentclass{epl}
\usepackage{graphicx}
\usepackage{amssymb}
\usepackage{amsmath}

\def\ba{\begin{eqnarray}}
\def\ea{\end{eqnarray}}
\def\beq{\begin{eqnarray}}
\def\eeq{\end{eqnarray}}
\def\be{\begin{equation}}
\def\ee{\end{equation}}

\title{Non Markovian persistence in the diluted Ising model at
  criticality} 
\shorttitle{Non Markovian persistence in the diluted Ising...} 
\author{Raja Paul \and Gr\'egory Schehr }
\institute{Theoretische Physik, Universit\"at des Saarlandes,
66041 Saarbr\"ucken, Germany}


\begin{document}

\maketitle

\begin{abstract}
We investigate global persistence properties for the non-equilibrium
critical dynamics of the randomly diluted Ising model. The disorder averaged 
persistence probability $\overline{{P}_c}(t)$ of the global
magnetization is found to 
decay algebraically with an exponent $\theta_c$ that we
compute analytically in a dimensional expansion in 
$d=4-\epsilon$. Corrections to Markov process are found to occur
already at one loop order and $\theta_c$ is thus a novel exponent
characterizing this disordered critical point.   
Our result is thoroughly compared with Monte Carlo simulations in
$d=3$, which also include a measurement of the initial slip
exponent. Taking carefully into 
account corrections to scaling, $\theta_c$ is found to be a universal
exponent, independent of the dilution factor $p$ along the critical
line at $T_c(p)$, and in 
good agreement with our one loop calculation.  
\end{abstract}

Persistence properties, and related first-passage-time problems, have been
intensively investigated both theoretically and experimentally these
last few years~\cite{satya_review}. Originally introduced in the context of
non-equilibrium coarsening dynamics of ferromagnets at zero temperature
\cite{persist_zero_T}, the persistence exponent $\theta$ describes the
long time decay ${P}(t) \sim t^{-\theta}$ of the probability ${P}(t)$
that a stochastic variable, here the local order 
parameter, does not cross a threshold value
during a time interval $t$. Interestingly, this exponent $\theta$
appeared to be 
highly non trivial, even for simple pure systems such as 
diffusion~\cite{majumdar_diffusion}, and therefore it has motivated numerous
investigations in a various pure models. However, 
except for few models in one
dimension~\cite{fisher_rfim,rieger_persist}, very little is known
about this exponent for disordered systems.  

Although for a ferromagnet at finite $T$, one expects
that the persistence ${P}(t)$ associated to the {\it local}
magnetization has an exponential tail,  
persistence properties received a special attention in the context
of critical dynamics of pure ferromagnets at $T_c$. 
Indeed, it has been proposed~\cite{majumdar_critical} that the 
probability ${P}_{\text{c}}(t)$ that the global magnetization
$M$ has not changed 
 sign in the time interval $t$ following a quench from a random
initial configuration, 
decays algebraically at large time ${P}_c(t) \sim
t^{-\theta_c}$. In this context, analytical progress is made
possible thanks to the 
property that, in the thermodynamic limit, the global order parameter
remains Gaussian at all finite 
times $t$. Indeed, for a $d$-dimensional system of linear size $L$,
$M(t)$ is the sum 
of $L^d$ random 
variables which are correlated only over a {\it finite} correlation length
$\xi(t)$. Thus, in the thermodynamic limit $L/\xi(t) \gg 1$, the
Central Limit Theorem (CLT) asserts that $M(t)$ is a Gaussian
process, for which powerful tools have been developed to compute the
persistence
properties~\cite{satya_review,satya_clement_persist}. Remarkably,
under the {\it additional} assumption  
that $M$ is a Markovian process, $\theta_c$ can be related to the
other critical exponents via the scaling relation $\theta_c =\mu
\equiv (\lambda - d + 1 - \eta/2)z^{-1}$, with $z$ and $\lambda$ the dynamical and
autocorrelation~\cite{janssen_rg, huse_lambda} exponent
respectively. However, as argued in 
Ref.~\cite{majumdar_critical}, $M$ is in general non Markovian and
thus $\theta_c$ is a {\it new} exponent associated to critical
dynamics. For the non conserved critical dynamics of
pure $\text{O}(N)$ model, corrections to this scaling relation were
indeed found to occur at two loops order~\cite{oerding_persist}, 
in rather good agreement with numerical simulations in dimensions
$d=2,~3$~\cite{Schulke97}. The question whether such 
a {\it universal} persistence exponent can also be defined, and
computed, in the presence of quenched disorder remains an open question. 

On the other hand, there is currently a wide interest in the slow
relaxational dynamics following a quench at a critical 
point~\cite{calabrese_review}.
Although simpler to study than glasses,
they display interesting properties such as two time dynamics
(aging) or violation of the Fluctuation Dissipation Theorem, as found
in more complex glassy
systems~\cite{cugliandolo_leshouches}. Recently, some  
progress has been achieved in the characterization of the effects of
the disorder on these 
properties~\cite{kissner_rim, calabrese_rim, schehr_rim_pre}.   
In this Letter, taking advantage of these recent studies,
we study persistence properties, both analytically using RG and numerically
via Monte Carlo simulations, of the randomly diluted Ising model:  
\begin{eqnarray}\label{def_diluted}
H = \sum_{\langle i j \rangle} \rho_i \rho_j s_i s_j
\label{eq_Hamil}
\end{eqnarray}  
where $s_i$ are Ising spins on a $d$-dimensional hypercubic lattice
and $\rho_i$ are {\it quenched} random variables such that
$\rho_i = 1$ with probability $p$ and
$0$ otherwise. For the experimentally relevant case of
dimension $d=3$~\cite{belanger_exp}, for which the specific heat
exponent of the pure 
model is positive, the disorder is expected, according to Harris
criterion\cite{harris_criterion}, to modify the universality class of the
transition. For $1-p \ll 1$,  the
large scale properties of (\ref{def_diluted}) at criticality
are then described by the
following $\text{O}(1)$ model with a random mass term, the so-called Random
Ising Model (RIM): 
\begin{equation}\label{H_rim}
H^{\psi}[\varphi] = \int d^d x \left[ \frac{1}{2}(\nabla \varphi)^2 +
 \frac{1}{2} [r_0 + \psi(x)] \varphi^2 + \frac{g_0}{4!}
 \varphi^4 \right]  
\end{equation} 
where $\varphi \equiv \varphi(x)$ and  
$\psi(x)$ is a Gaussian random variable $\overline{\psi(x)
  \psi(x')} = \Delta_0 \delta^d(x-x')$ and $r_0$, the bare mass, is
  adjusted so that the renormalized one is zero. 

We first study the relaxational dynamics of the Randomly diluted Ising
Model (\ref{H_rim}) in dimension $d=4-\epsilon$ described by a Langevin equation:
\begin{eqnarray}\label{def_Langevin}
\eta \frac{\partial}{\partial t} \varphi(x,t) = - \frac{\delta
 H^{\psi}[\varphi]}{\delta \varphi(x,t)} + \zeta(x,t)
\end{eqnarray}
where $\langle \zeta(x,t) \rangle = 0$ and
$\langle \zeta(x,t) \zeta(x',t') \rangle = 2\eta T \delta(x-x')
\delta(t-t') $ is the thermal noise and $\eta$ the friction
coefficient, set to 
$1$ in the following. At initial time $t_i = 0$, the system is in a random
configuration with zero mean magnetization and short range
correlations $[\varphi(x,t=0) \varphi(y,t=0)]_i = \tau_0^{-1} \delta^d (x-y)$, 
where $\tau_0^{-1}$ is irrelevant in the RG sense~\cite{janssen_rg} (it will thus be set 
to zero in the following). Defining the global magnetization $M(t)$ as
\begin{eqnarray}
M(t) = \frac{1}{N_{\text{occ}}} \sum_i \rho_i s_i(t) = \frac{1}{L^d}\int_x
\varphi(x,t) \label{def_mag}
\end{eqnarray}
where $N_{\text{occ}}$ is the total number of occupied sites, 
we are interested in the disorder averaged probability
$\overline{{P}_c}(t)$ 
that the magnetization has 
not changed sign in the time interval $t$ following the quench.
In that purpose, we introduce the correlation ${\cal
C}_{tt_w} = \overline{\langle M(t)
      M(t_w)  \rangle}$ 
and the linear response ${\cal R}_{tt_w}$ to a small external
field ${f}(t_w)$,  ${\cal R}_{tt_w} = \overline{{\delta \langle
  M(t) \rangle}/{\delta 
  {f}(t_w)} }$ 
where $\overline{...}$ and $\langle ... \rangle $
denote respectively averages over the disorder and 
thermal fluctuations. At one loop order, ${\cal C}_{tt_w}$ and ${\cal
  R}_{tt_w}$ take the scaling forms~\cite{calabrese_rim}, ${\cal C}_{t
  t_w} = A_c t^{\frac{2-\eta}{z}} \Phi_c({t}/{t_w})$ and ${\cal
  R}_{t t_w} = A_r t^{\frac{2-\eta-z}{z}} \Phi_r({t}/{t_w})$ 
where $A_c, A_r$ are non universal constants and, $\Phi_{c,r}(x)$ -- not shown
here for clarity  -- are universal scaling functions such that
$\Phi_{c,r}(1)=1$.  
In a previous publication~\cite{schehr_rim_pre}, we have shown, using
the Exact Renormalization Group for the dynamical effective
action~\cite{schehr_co_pre}, that they are given, at this order, by
the solution of the following equations
\begin{eqnarray}
&&\partial_{t} {{\cal R}}_{tt_w} + \mu(t)
{{\cal R}}_{tt_w} +
\int_{t_w}^t d {t_1} \Sigma_{tt_1}
{{\cal R}}_{t_1t_w} = 0 \quad, \quad  \mu(t) = - \int_{t_i}^t d {t_1}
\Sigma_{tt_1}  \label{Eq_R} \\ 
&&{\cal C}^{}_{ {t} {t_w}} =
2T\int_{t_i}^{{t_w}} dt_1 {{\cal R}}^{}_{ {t}t_1}{{\cal
R}}^{}_{ {t_w}t_1} 
+ \int_{t_i}^{ {t}} dt_1 \int_{t_i}^{ {t_w}} dt_2{{\cal
R}}^{}_{ {t} t_1} D_{t_1t_2}
{{\cal R}}^{}_{ {t_w} t_2}   
\label{Eq_C}
\end{eqnarray}
where the self energy $\Sigma_{t_1 t_2}$ and the noise-disorder kernel 
$D_{t_1 t_2}$ have been computed in~\cite{schehr_rim_pre} :
\begin{eqnarray}
&& \Sigma_{t_1 t_2} = -\frac{1}{2} \sqrt{\frac{6\epsilon}{53}} 
  (\gamma(t_1-t_2))^2 \quad, \quad D_{t_1 t_2} = \frac{T_c}{2}
  \sqrt{\frac{6\epsilon}{53}} 
  \left( \gamma(t_1-t_2) - \gamma(t_1+t_2)\right)
\end{eqnarray} 
where $\gamma(x) = (x+\Lambda_0^{-2})^{-1}$, $\Lambda_0$ being the UV
cutoff~\cite{foot_UVcutoff}. These equations Eq.~(\ref{Eq_R}, 
\ref{Eq_C}), suggest 
to describe the time evolution of the magnetization by an
effective Gaussian process $\tilde M(t)$: 
\begin{equation}
\partial_t \tilde M(t) + \mu(t) \tilde M(t) = -\int_{0}^{t} dt_1
\Sigma_{tt_1} \tilde M(t_1) + {\tilde \zeta}(t) \label{Eff_Eq}
\end{equation}
where $\tilde \zeta(t)$ is an effective 
disorder induced Gaussian noise with zero mean and correlations
$\langle \tilde \zeta(t)\tilde \zeta(t')\rangle_{\text{eff}}  = 2
T\eta \delta(t-t') 
+ D_{tt'}$. The idea is then to compute 
$\overline{{P}_c}(t)$ as the persistence probability of the
process $\tilde 
M(t)$. The rhs of Eq.~(\ref{Eff_Eq}) clearly indicates
that this process is non-Markovian. However, as $\Sigma_{tt'}$ as well
as $D_{tt'}$ are of order 
${\cal  O}({\sqrt{\epsilon}})$, one can use the perturbative computation of
$\theta_c$ around a Markov process initially developed in
Ref.~\cite{satya_clement_persist} and further studied in the context
of critical dynamics in Ref.~\cite{oerding_persist}. In that
purpose~\cite{satya_review}, let us
introduce the normalized Gaussian process ${m}(t) = {\tilde
  M(t)}/\sqrt{{\langle \tilde M^2(t)\rangle_{\text{eff}}}}$. Let 
$T=\ln{t}$, then ${m}(T)$ is a {\it stationary} Gaussian process, its
persistence properties are then obtained from its autocorrelation function:
\begin{eqnarray}
&&\langle{m}(T) {m}(T_w)\rangle_{\text{eff}} = e^{-\mu(T-T_w)}{\cal
      A}(e^{T-T_w}) \label{TTI} \\ 
&& {\cal A}(x) = \left(1 +
  \frac{1}{4}\sqrt{\frac{6\epsilon}{53}}(x \log{\frac{x-1}{x+1}} -
  \log{\frac{x^2-1}{4 x^2}}    )  \right) \quad, \quad \mu
      =\frac{\lambda-d+1-\eta/2}{z}  \nonumber 
\end{eqnarray}     
Under this form (\ref{TTI}), one can use the first order perturbation
theory result of Ref.~\cite{oerding_persist} to obtain the one loop
estimate:   
\begin{eqnarray}
\Delta \equiv  \theta_c - \mu = \sqrt{\frac{6
    \epsilon}{53}} \frac{\sqrt{2}-1}{2} 
+ {\cal O}(\epsilon) = 0.06968... \quad \text{in} \quad d=3
    \label{persist_one_loop} 
\end{eqnarray}
where $\mu$ is the value corresponding to a Markov
process. The second term in Eq.~(\ref{persist_one_loop}) is the first
correction due to the non-Markovian
nature of the dynamics. Interestingly, this correction is entirely
determined by the non-trivial structure of the scaling function ${\cal A}(x)$,
which is directly related to $\Phi_c(x)$. Notice also that, at variance with
the pure $\text{O}(N)$ models 
\cite{majumdar_critical, oerding_persist}, these corrections in
presence of quenched static disorder already
appear at one loop order.  

We now turn to the results from our Monte Carlo simulations of the
relaxational dynamics of the randomly diluted Ising model
(\ref{def_diluted}) in dimension $d=3$, which were done on $L^3$ 
cubic lattices with periodic boundary conditions. 
For $p=1$, the system is pure and shows ferromagnetic order at $T<T_c(p=1)$.
The critical temperature $T_c(p)$~\cite{ballesteros_tc,heuer_tc}
decreases with $p$ and becomes 0 at the percolation
threshold $p=p_c$ (for a $d=3$
cubic lattice, $p_c \simeq 0.341$). For $p>p_c$, the system 
is initially prepared in a random initial configuration with zero mean
magnetization $M_0 = 0$. Up and down spins are randomly 
distributed on the occupied sites, mimicking a high-temperature
disordered configuration before the quench.
At each time step, one site is randomly chosen and the move 
$s_i \to - s_i$ is accepted or
rejected according to Metropolis rule. One time unit corresponds to
$L^3$ such time steps.        
The exponent $\theta_c$ is measured numerically for cubic lattices of linear
size $L =$ 8, 16, 32 and 64. After a quench to $T_c$ from the initial
random configuration each system evolves until the global
magnetization first change sign. $\overline{P}_c(t)$ is then measured
as the fraction of surviving systems at each time $t$, over a
number of samples which 
varies from $2 \times 10^5$ for $L=8$ to $2 \times 10^4$ for
$L=64$. Although the Renormalization Group analysis stays that the
critical exponents are independent of $p$ (at least for $1-p \ll 1$)
the question of universality in disordered systems, and in particular in
the present problem~\cite{heuer_tc}, is a 
longstanding issue. Therefore we will compute 
$\overline{P}_c(t)$ for different values of 
$p(>p_c)=$ 0.499, 0.6, 0.65 and 0.8 along the critical line at $T_c(p)$.  
 
\begin{figure}
\onefigure[scale=0.3]{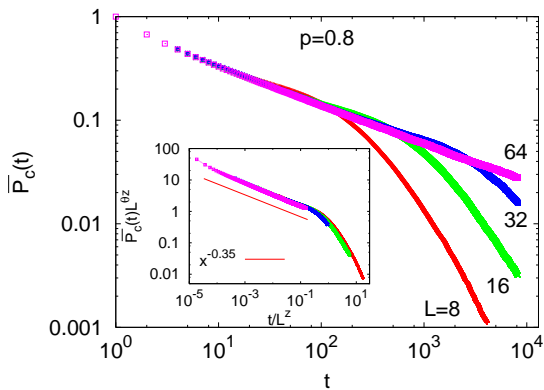}
\caption{Persistence probability $\overline{P_c}(t)$ plotted in the
  log-log scale for $p=0.8$ and different $L$, with $M_0=0$. {\bf
    Inset:} Scaled plot of 
  $L^{\theta_c z} \overline{P_c}(t)$ vs $t/L^z$ for different $L$
with $z=2.62$ and $\theta = 0.35$.}
\label{fig1}
\end{figure}

\begin{figure}
\twoimages[scale=0.28]{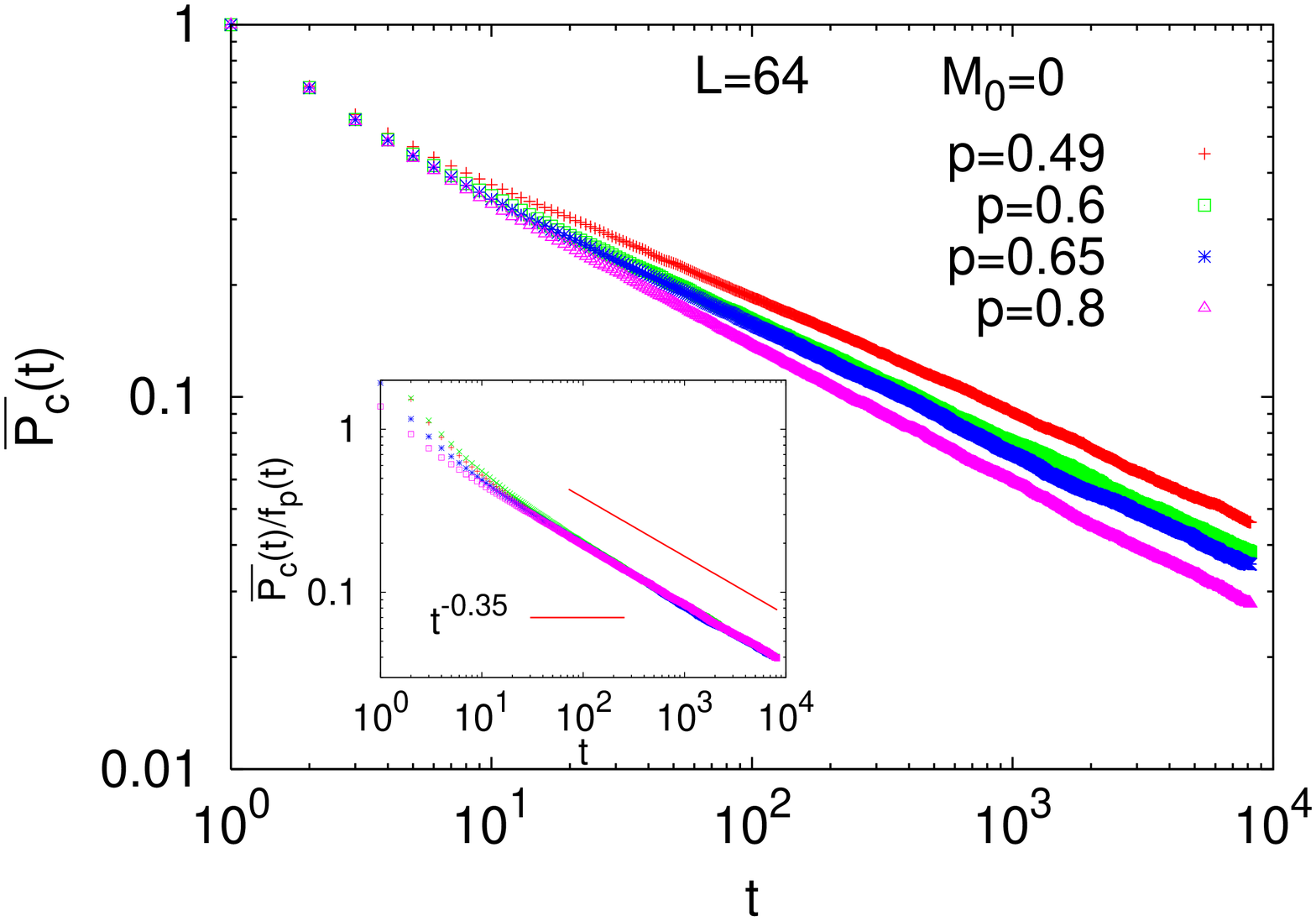}{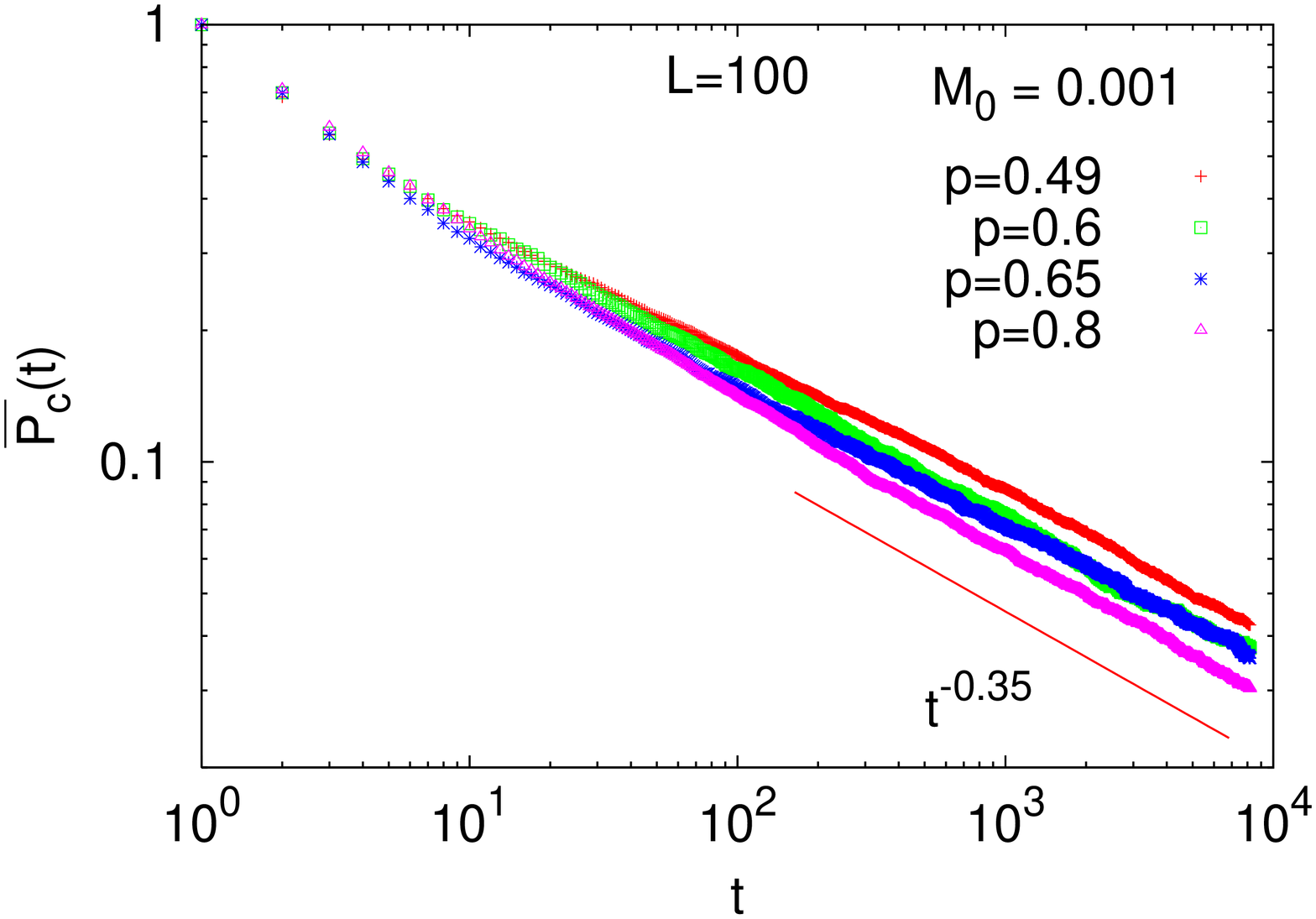}
\caption{{\bf Left:} Persistence probability $\overline{P_c}(t)$ for different
  $p=$ 0.5, 0.6, 0.65 and 0.8 with fixed $L=64$. The system is initially
  prepared with $M_0=0$. {\bf Inset:} Universality of $\overline{P_c}(t)$ for
  different $p$. The function $f_p(t)$ is defined in the
  Eq.~(\ref{persit_correc_scaling}). {\bf Right:} Similar to
  fig.~\ref{fig2} (left), but prepared with a non-zero 
  initial magnetization $M_0=0.001$. The linear system size in this
  case is $L=100$.}
\label{fig2}
\end{figure}

In fig.~\ref{fig1} we present the results of ${\overline P}_c(t)$ for
$p=0.8$ and for 
different lattice sizes. According to standard finite-size
scaling~\cite{majumdar_critical}, one expects the scaling form
$\overline{P}_c(t) = t^{-\theta_c} f(t/L^z)$,
where $z$ is the dynamical exponent. Keeping the rather well
established value of
$z=2.62(7)$~\cite{parisi_simu_rim, schehr_rim_pre} fixed, $\theta_c$
is varied to 
obtain the best data collapse. The final scaled plot is shown in
the inset of  
fig.~\ref{fig1}. We have checked (not shown) that the
same analysis can be done for all other values of $p$ studied
here. However, the best data collapse is then obtained for a value
of $\theta_c$ which is $p$-dependent. This is depicted in
  fig.~\ref{fig2} (left) where one clearly observes a decrease in slope
while occupation probability $p$ is reduced, which could naively
suggest a non universal value for the exponent $\theta_c(p)$.
However, it is now well known that the present model is strongly
affected by corrections to
scaling~\cite{parisi_simu_rim,schehr_rim_pre}, characterized by a
universal exponent $b$, {\it i.e.} 
independent of $p$, which has been
estimated in previous works~\cite{parisi_simu_rim, schehr_rim_pre} to
be $b = 0.23(2)$. Therefore following these previous 
studies we suggest that the persistence can be written as
\begin{eqnarray}
\overline{P_c}(t) = t^{-\theta_c} f_p(t) \quad, \quad f_p(t) = A(p) (1
+ B(p) t^{-b}) \label{persit_correc_scaling}
\end{eqnarray}
which $A(p)$ and $B(p)$ are fitting parameters. As shown in the inset of
fig.~\ref{fig2} (left), the good collapse of the quantity
$P(t)/f_p(t)$ for different values of $p$ suggest that $\theta_c$ is a
{\it universal} exponent. $A(p)$ and $B(p)$ are found to be monotonous
decreasing function of $p$ and such that $B(p=0.8) \simeq 0$.
The value extracted from this collapse is
\begin{eqnarray}
\theta_c = 0.35 \pm 0.01 \label{num_value}
\end{eqnarray}

We have also computed the persistence probability for systems quenched from 
random configuration with a small initial magnetization
$M_0=0.001$. The number of up $N_{\text{up}}$ and down
$N_{\text{down}}$ spins are 
thus : $N_{\text{up}} = (1+M_0)/{2N_{\text{occ}}}$ and $N_{\text{down}} =
N_{\text{occ}}-N_{\text{up}}$. First we randomly distribute the
$N_{\text{up}}$ up spins in 
the occupied sites of the lattice and then fill up the rest with down spins.
As noticed previously~\cite{Schulke97} this protocol
allows to reduce the statistical noise and thus to study larger system
sizes (this however renders the finite size scaling analysis more
subtle~\cite{Schulke97}). In fig.~\ref{fig2} (right), we plot the persistence
for system size $L=100$ ( the data have been averaged over $2\times 10^4$
ensembles) for $p=$ 0.499, 0.6, 0.65 and 0.8. The study of larger
system size allows to reduce the corrections to scaling. Indeed, 
although in the short time scales, the
straight lines have slightly different slopes which depends upon
$p$, at later times the slopes varies from $0.36(1)$ for
$p=0.8$ to $0.35(1)$ for $p=0.499$ :  this confirms the
$p$-independent value of $\theta_c$ obtained previously
Eq.~(\ref{num_value}).    

In order to compare this numerical value (\ref{num_value}) with our
one loop calculation (\ref{persist_one_loop}) one needs an estimate
for $\mu$. Such an estimate is needed
not only for the sake of this comparison but also to characterize
quantitatively non Markovian effects. The argument mentioned in
the introduction, relying on the CLT, which stays
that the global magnetization is, in the thermodynamic limit, a
Gaussian variable is also valid in the presence of disorder, and we
have checked it numerically. Therefore, a
finite difference $\Delta$ (\ref{persist_one_loop}) is the signature
of a Non Markovian 
process. A numerical estimate of $\mu$ can be done in two different
ways~\cite{mu_rg_estimate}. First, using the numerical estimates for
$\eta = 0.0374(45)$ from 
Ref.~\cite{ballesteros_tc}, $z = 2.62(7)$ from
Ref.~\cite{parisi_simu_rim}, and  $\lambda/z
= 1.05(3)$ from Ref.~\cite{schehr_rim_pre} one obtains $\mu_{\text{num}}
= 0.28(4)$, which gives a first estimate
$\Delta_{\text{num}} = 0.07(5)$. Because of the relatively big
error bars on $\lambda$ and $z$, we propose, 
alternatively, to express $\mu$ in terms of the initial slip exponent
, $\theta'$ 
\cite{janssen_rg}, using
$\lambda= d-z\theta'$, as $\mu = -\theta' + (1-\eta/2)/z$. Although
$\theta'$ has been estimated numerically for pure Ising systems
\cite{LRZ94,Grassberger94}, there are no available data for the
disordered case : we thus now present a numerical computation of
$\theta'$. 

\begin{figure}
\onefigure[scale=0.3]{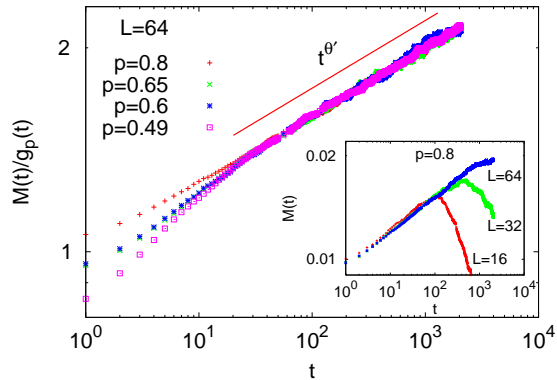}
\caption{Rescaled global magnetization $M(t)/g_p(t)$ as a function
of $t$ for $p=$ 0.49, 0.6, 0.65 and 0.8 in the log-log scale. The
linear system size is $L=64$, the initial magnetization is $M_0 =
0.01$ and the measured exponent $\theta'=0.1$. {\bf Inset:}
Global magnetization $M(t)$ for $p=0.8$ and $L=16, 32$ and $64$.}
\label{fig_initial_slip}
\end{figure}

In that purpose we study the time evolution of $M(t)$ 
when the system is quenched from an initial configuration with short
range correlations but a finite, however small,  magnetization
$M_0$. The initial stage 
of the dynamics is characterized by an 
increase of the global magnetization, described by a universal power
law~\cite{janssen_rg}
\begin{equation}
M(t) \sim M_0 t^{\theta'} \label{eq_ini_slip}
\end{equation}
At larger times, $t\gg t_0$, $M(t)$ decreases to zero as $M(t) \sim
t^{-\beta/(\nu z)}$. In our simulations, the system is initially
prepared in a random 
initial configuration with a small but finite initial
magnetization $M_0 = 0.01$ and quenched down to $T_c$ at $t=0$. In
all subsequent times we measure $M(t)$ (\ref{def_mag})
for linear system sizes $L=$ 8, 16, 32 and 64. Finally data are
averaged over $8 \times 10^5$ samples for $L=8$ to $10^4$ samples for
$L=64$. In the 
inset of fig.~\ref{fig_initial_slip}, we show a plot of $M(t)$ for
$p=0.8$ and different system sizes. One sees clearly that $M(t)$ is
increasing until a time $t_0$, which is an increasing
function of the 
system sizes considered here~\cite{foot_t0}, and then decreases to
zero (although the 
aforementioned scaling for $t \gg t_0$ is not clearly seen here). Here also,
we compute $M(t)$ for different values of $p=0.499, 0.6, 0.65$
and $0.8$ and observe corrections to scaling which we take into
account as $M(t) = t^{\theta'} g_p(t)$ with
$g_p(t)=A'(p)[1+B'(p)t^{-b}]$, $A'(p),~B'(p)$ being fitting parameters. 
As shown in
fig.~\ref{fig_initial_slip}, one obtains a reasonably good data
collapse of $M(t)/g_p(t)$ 
vs.~$t$ for the different values of $p$. Here also, one finds that
$p=0.8$ is almost unaffected by these corrections to scaling, {\it
  i.e.} $B'(0.8) \simeq 0$. After a microscopic time
scale, which increases as $p$ is lowered, one observes a universal
power law increase (\ref{eq_ini_slip}), from which we get the estimate
$\theta'=0.10(2)$. This value is in good agreement with the
two-loops estimate $\theta'_{\text{2 loops}} = 0.0868$
\cite{oerding_randommass_theta_twoloops}
and agrees also quite well with the scaling relation $\lambda = d -
z\theta'$~\cite{schehr_rim_pre}. This gives $\mu_{\text{num}} =
0.27(3)$ and our final numerical estimate 
\begin{eqnarray}
\Delta_{\text{num}} = 0.08 \pm 0.04 
\label{estim_delta}
\end{eqnarray} 
which is in good agreement with our previous one loop estimate
(\ref{persist_one_loop}). We 
also notice that these deviations from a Markov process are
slightly larger than for the pure case~\cite{Schulke97}.

In summary we have shown that the global persistence
exponent $\theta_c$ carries the signature of Non Markovian effects,
and is thus a new critical exponent characterizing this random
critical point, that we have computed within a one loop approximation
(\ref{persist_one_loop}). Our detailed numerical analysis, which
relies upon a computation of $\theta_c$ and the initial slip exponent
$\theta'$, supports the existence of these
Non-Markovian violations (\ref{estim_delta}) and is in good
quantitative agreement with our perturbative approach. In addition,
$\theta_c$ is found to be universal along the critical line.      
In view of recent progress in the study of aging
properties in finite dimensional glassy phases~\cite{schehrledou}, it
would be very  
interesting to extend the approach presented
here to these situations.

\acknowledgments

GS acknowledges the financial support provided
through the European Community's Human Potential Program 
under contract HPRN-CT-2002-00307, DYGLAGEMEM and RP's work was
supported by the DFG (SFB277).


\begin{thebibliography}{99}


\bibitem{satya_review}
  For a review, see 
  \Name{Majumdar S. N.}
  \REVIEW{Curr. Sci.} {77} {1999} {370}.

\bibitem{persist_zero_T}
  \Name{Derrida B., Bray A. J. \and Godr$\grave{\text{e}}$che C.}
  \REVIEW{J. Phys. A} {27} {1994} {L357}.

\bibitem{majumdar_diffusion}
  \Name{Majumdar S. N., Sire C., Bray A. J. \and Cornell S. J.} 
  \REVIEW{Phys. Rev. Lett.} {77} {1996} {2867}.

\bibitem{fisher_rfim}
  \Name{D.S. Fisher, P. Le Dousssal and C. Monthus}
  \REVIEW{Phys. Rev. Lett.} {80} {1998} {3539}. 


\bibitem{rieger_persist}
  \Name{Rieger H. and Igloi F.}
  \REVIEW{Europhys. Lett.} {45} {1999} {673}. 



\bibitem{majumdar_critical}
  \Name{Majumdar S. N, Bray A. J., Cornell S. J. \and Sire C.}
  \REVIEW{Phys. Rev. Lett.} {77} {1996} {3704}. 


\bibitem{satya_clement_persist}
  \Name{Majumdar S. N. \and Sire C.}
  \REVIEW{Phys. Rev. Lett.} {77} {1996} {1420}.


\bibitem{janssen_rg} 
	\Name{Janssen H. K., Schaub B. \and Schmittmann B.}
	\REVIEW{Z. Phys. B} {73} {1989} {539}. 


\bibitem{huse_lambda}
  \Name{Huse D. A.}
  \REVIEW{Phys. Rev. B}{40} {1989}{304}.

\bibitem{oerding_persist}
  \Name{Oerding K., Cornell S. J. \and Bray A. J.}
  \REVIEW{Phys. Rev. E} {56} {1997} {R25}.


\bibitem{Schulke97}
  \Name{Stauffer D.}
  \REVIEW{Int. J. Mod. Phys. C}{7}{1996}{753};
  \Name{Sch\"ulke L. \and Zheng B.}
  \REVIEW{Phys. Lett. A} {233} {1997} {93}.



\bibitem{calabrese_review}
  \Name{Calabrese P. \and Gambassi A.}
  \REVIEW{J. Phys. A} {38} 2005{} {R133-R193}.



\bibitem{cugliandolo_leshouches}
For a review see  \Name{Cugliandolo L. F.}
  \Book{Dynamics of glassy systems, in {\it Slow relaxation
      and nonequilibrium dynamics in condensed matter}, J.~L.~Barrat
    {\it et al.}}, 
  Springer-Verlag, 2002. 


\bibitem{kissner_rim}
\Name{Kissner J. G.}
\REVIEW{Phys. Rev. B} {46} {1992} {2676}.

\bibitem{calabrese_rim}
  \Name{Calabrese P. \and Gambassi A.}
  \REVIEW{Phys. Rev. B} {66} {2002} {212407}.


\bibitem{schehr_rim_pre}
  \Name{ Schehr G. \and Paul R.} 
  \REVIEW{Phys. Rev. E} {72} {2005} {016105}.
  



\bibitem{belanger_exp}
  \Name{Belanger D. P.}
  \REVIEW{Braz. Journ. Phys.} {30(4)} {2000} {682}.
 

\bibitem{harris_criterion}
  \Name{Harris A. B.}
  \REVIEW{J. Phys. C} {7} {1974} {1671}.



\bibitem{schehr_co_pre}
  \Name{Schehr G. \and Le Doussal P.}
  \REVIEW{ Phys. Rev. E} {68} {2003} {046101}.




\bibitem{foot_UVcutoff}
  One chooses here a regularization of
  integrals over Fourier modes $q$ of the form $\int_q \equiv \int_q
  e^{-q^2/2\Lambda_0^2}$.   



\bibitem{heuer_tc}
  \Name{Heuer H.-O.}
  \REVIEW{J. Phys. A: Math. Gen.}{26}{1993}{L333};
  \SAME{26}{1993}{L341}.

\bibitem{ballesteros_tc}
  \Name{Ballesteros H. G., Fern\`andez L. A., Mart\'in-Mayor V. \and
    Munoz~Sudupe A.} 
  \REVIEW{Phys. Rev. B}{58}{1998}{2740}.

\bibitem{parisi_simu_rim}
  \Name{Parisi G., Ricci-Tersenghi F. \and Ruiz-Lorenzo J. J.}
  \REVIEW{Phys. Rev. E} {60} {1999} {5198}.

\bibitem{mu_rg_estimate}
The perturbative expansion, using RG, of $\mu$ gives :
$\mu_{1 loop} = 0.415$ at one loop order and $\mu_{2 loops} = 0.357$
at two loops. As this loop expansion gives rather
poor estimate for some exponents entering the expression of $\mu$,
{\it e.g.} $z$~\cite{parisi_simu_rim}, we prefer to rely upon a
numerical estimate of it.   



\bibitem{LRZ94}
  \Name{Li Z.-B., Ritschel U. \and Zheng B.}
  \REVIEW{J. Phys. A: Math. Gen.} {27} {1994} {L837}.


\bibitem{Grassberger94}
  \Name{Grassberger P.}
  \REVIEW{Physica A} {214} {1995} {547}.

\bibitem{foot_t0}
In the thermodynamic limit and $M_0 \to 0$, it is expected to scale as
$t_0 \sim M_0^{-1/(\theta'+\beta/(\nu z))}$~\cite{janssen_rg}.  




\bibitem{oerding_randommass_theta_twoloops}
\Name{Oerding K. \and Janssen H.K.}
\REVIEW{J. Phys. A} {28} {1995} {4271}.



\bibitem{schehrledou}
\Name{Schehr G. \and Le Doussal P.}
\REVIEW{Phys. Rev. Lett.} {93}{2004}{217201};
\REVIEW{Europhys. Lett.}{71}{2005}{290}. 


\end{thebibliography}
\end{document}